\documentclass[useAMS,usenatbib]{mn2e}

\usepackage{psfig,graphicx}

\input epsf
\input psfig.sty
\usepackage{amsmath}
\usepackage{amssymb}
\usepackage{upgreek}
\usepackage{graphicx}
\usepackage{longtable}

\newcommand{\arcs}{\mbox{\ensuremath{^{\prime\prime}}}}

\title[Near-IR spectra of Galactic AGB stars]
  {Near-IR spectra of IPHAS extremely red Galactic AGB stars}
\author[N.J.~Wright et al.]
  {N.J.~Wright,$^{1,2}$ M.J.~Barlow,$^2$ R.~Greimel,$^{3}$ J.E.~Drew,$^{4}$ M.~Matsuura,$^{5,6}$
  \newauthor Y.C.~Unruh$^7$ and A.A.~Zijlstra$^8$\\
  \\
  $^1$Harvard-Smithsonian Center for Astrophysics, 60 Garden Street, Cambridge, MA~02138, U.S.A.\\
  $^2$Department of Physics and Astronomy, University College London, Gower Street, London WC1E 
6BT, U.K.\\
  $^3$Institut f\"ur Physik, Karl-Franzen Universit\"at Graz, Universit\"atsplatz 5, 8010 Graz, Austria\\
  $^4$Centre for Astrophysics Research, University of Hertfordshire, College Lane, Hatfield, AL10 9AB, 
U.K.\\
  $^5$UCL Institute of Origins, Department of Physics and Astronomy, University College London, Gower Street, London, WC1E~6BT, U.K.\\
  $^6$UCL Institute of Origins, Mullard Space Science Laboratory, University College London, Holmbury, St.~Mary, Dorking, Surrey, RH5~6NT, U.K.\\
  $^7$Imperial College of Science, Technology and Medicine, Blackett Laboratory, Exhibition Road, London, SW7 2AZ, U.K.\\
  $^8$Jodrell Bank Centre for Astrophysics, The University of Manchester, School of Physics and Astronomy, Manchester, M13 9PL, U.K.\\
  }

\pagerange{\pageref{firstpage}--\pageref{lastpage}} \pubyear{2009}

\def\LaTeX{L\kern-.36em\raise.3ex\hbox{a}\kern-.15em
    T\kern-.1667em\lower.7ex\hbox{E}\kern-.125emX}

\begin{document}

\label{firstpage}

\maketitle

\begin{abstract}

We present a library of 139 near-IR spectra of cool asymptotic giant branch stars that will be useful for comparison with theoretical model atmosphere calculations and for modeling the integrated emission from intermediate-age stellar populations. 
The source list was selected from the ``extremely red'' region of the INT Photometric H$\alpha$ Survey (IPHAS) colour-colour plane that is overwhelmingly dominated by very late-type stars. The spectral library also includes a large fraction of S-type and carbon stars. 
We present a number of spectral classification sequences highlighting the various molecular features identified and discuss a number of rare features with uncertain identifications in the literature. 
With its focus on particularly cool photospheres this catalogue serves as a companion to recent spectroscopic atlases of MK standards in the near-IR.
Finally the relationship between IPHAS $(r' - i')$ and $(r' - $H$\alpha)$ colours and spectroscopically determined properties is discussed and a strong correlation between $(r' - $H$\alpha)$ colour and the C/O abundance index for S-type and carbon stars is noted. This relation has the potential to separate O-rich, S-type and carbon stars in the Galaxy based on their photometry alone.

\end{abstract}

\begin{keywords}
stars: AGB and post-AGB - stars: chemically peculiar - stars: carbon - infrared: stars - atlases - techniques: spectroscopic
\end{keywords}

\section{Introduction}

Asymptotic giant branch (AGB) stars represent one of the last evolutionary stages passed through by all intermediate mass stars ($0.8 <$~M/M$_{\odot} < 8.0$) and are responsible for large amounts of processed material returned to the interstellar medium (ISM). Their cool extended atmospheres and winds are prime sites for the development of molecular chemistry and the formation of dust grains. AGB stars are also some of the most luminous stars in a galaxy and are the most dominant source of near-IR light from intermediate age ($10^8 - 10^9$ yr old) stellar populations \citep{lanc00}.  In particular, the reddest, thermally pulsating AGB stars may contribute as much as 80\% of the integrated population light of a galaxy in the $K$-band \citep{lanc00}.

The spectral energy distributions of AGB stars peak in the near-IR, which also offers some of the most prominent molecular features that are sensitive to both surface gravity and effective temperature (e.g. TiO, H$_2$O, and CO). Improvements in infrared array detectors have led to considerable advances in the construction of infrared spectrographs, which have also led to developments in our understanding of the near-IR spectral region \citep[e.g.][]{joyc98,meye98,wall97}. Despite this, the majority of existing near-IR spectral libraries of AGB stars are limited to K and early-M types and contain few stars of late M-type. This is a product of both their relative local rarity and the high obscuration from circumstellar material produced during the evolution to the OH/IR phase.

This paper presents a library of over 100 near-IR spectra of AGB stars, including many S-type and carbon stars. To overcome the bias towards early-type M giants in existing spectral libraries, this library has a particular focus on late-type M giants, with more than half the classified O-rich sources of type M6 or later. The spectral resolution in the near-IR is sufficient to identify the majority of molecular bands and the strongest individual metal lines. This library will therefore be of use to those modelling the integrated emission from intermediate-age populations, those studying the evolution of very late-type AGB stars and for comparison with theoretical models of the spectra of late-type stars.


In Section~2 we discuss the selection of targets and the observational and data reduction process. In Section~3 we present the spectra, arranged by chemical type, spectral type and observational band and outline the methods used to perform spectral classification. Finally in Section~4 we present the results of the spectral classification, with particular reference to the optical and near-IR colours from which the targets were selected.

\section{Observations}

\begin{figure}
\begin{center}
\includegraphics[width=185pt, angle=270]{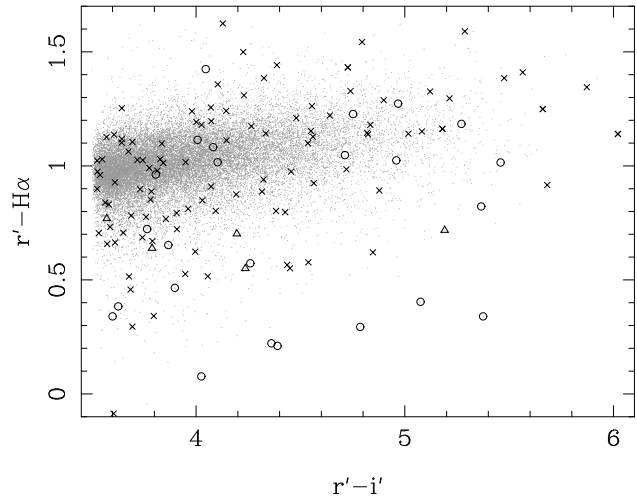}
\caption{IPHAS colour-colour diagram of the ``extremely red'' region showing the positions of all the extremely red stellar objects (ERSOs) as dots and the sources observed as symbols: cross symbols are O-rich sources, circles are S-type stars, triangles are carbon stars and asterisks are spectrally unidentified sources.}
\label{colcol_liris}
\end{center}
\end{figure}

The recent generation of deep photometric surveys of the Galactic Plane, such as the Isaac Newton Telescope (INT) Photometric H$\alpha$ Survey \citep[IPHAS,][]{drew05,gonz08} are greatly increasing the number of known objects in short-lived evolutionary phases in our Galaxy, including many AGB stars. While AGB stars are easily detected in the near-IR where the majority of their light is emitted, deep optical surveys offer other considerable advantages. In particular, the filter combination employed by IPHAS (Sloan $r'$, $i'$, and narrow-band H$\alpha$ filters) provides a colour-colour diagram ($r' - $H$\alpha$ versus $r' - i'$, e.g. Figure~\ref{colcol_liris}) where the dwarf and giant branches are very clearly separated \citep[e.g.][]{drew05}. 
Though AGB stars are fainter and suffer more obscuration in the visual than the near-IR, the high photometric depth of IPHAS ($r' = 20$ at 10$\sigma$), allows it to detect AGB stars at distances of several kpc and through extinctions up to $A_V \sim 10$ \citep[see Figure~5 of][]{wrig08}. 
For example, an AGB star of temperature class M, seen through a reddening of $E(B-V) \sim 2$~(1) will be detectable to a distance of $\sim$13~(45)~kpc. 
Given the survey's coverage of the entire northern Galactic Plane ($l = 30$~--~210, $-5^{\circ} \le b \le +5^{\circ}$) and the typical levels of extinction found outside the Solar circle ($E(B-V) \sim$~1~--~2), this will allow the great majority of AGB stars in this region to be detected. 
\citet{wrig08} studied a population of ``extremely red stellar objects'' (ERSO) identified from IPHAS data and classified as stellar sources with ($r' - i') > 3.5$. They showed that this region consisted almost entirely of highly reddened AGB stars, many with significant amounts of circumstellar material indicative of mass-loss rates appropriate for the tip of the AGB. 

Targets for spectroscopy were chosen to fully explore the entire ERSO region of the IPHAS colour-colour plane. We also attempted to obtain a good representation of sources across the Galactic plane, though this was partly limited by our observations taking place in the summer months at the Roque de Los Muchachos Observatory. 

Spectra of 139 sources were obtained using LIRIS \citep[Long-slit Intermediate Resolution Infrared Spectrograph,][]{acos03,manc04}, a near-IR camera/spectrometer on the 4.2 meter William Herschel Telescope (WHT) at the Roque de Los Muchachos Observatory in La Palma, Spain. 
Observations were performed on five nights in the summers of 2006 and 2007. Conditions were variable with thin clouds or high cirrus on three out of five nights. A list of the targets observed can be found in Table~\ref{observation_list} and their positions in the IPHAS colour-colour plane, relative to the ERSO population are shown in Figure~\ref{colcol_liris}. Objects with previous spectral classifications are listed in Table~\ref{appendix}.

\subsection{Observational procedure and data reduction}

Each object was observed using both the lrzj8 (0.89-1.51~$\mu$m) and lrhk (1.40-2.39~$\mu$m) grisms, giving continuous coverage from 0.89-2.39~$\mu$m through the z, J, H and K bands. 
The resolving power was 700 for both grisms, allowing many molecular features and the strongest metal lines to be identified. The spatial scale was 0.25\arcs~pixel$^{-1}$, and the slit width used during the observations was 1\arcs, aligned along the parallactic angle. Observations were performed using an ``ABBA'' telescope-nodding pattern, placing the source in two positions along the slit, A and B, separated by 80\arcs. Exposure times for each grism are listed in Table~\ref{observation_list}. The bias and dark levels of most near-IR detectors are unstable with time, therefore LIRIS takes a ``pre-read'' image, which is automatically subtracted from the ``post-read'' exposure (also known as ``double-correlated sampling'').

\newpage
\onecolumn
\footnotesize
\begin{longtable}{llrrrccll}
\caption[]{Sources observed with LIRIS on the WHT. Mean photometric magnitudes are taken from all IPHAS measurements. Spectral types for all sources are also listed (only the temperature class is listed since all sources are believed to be giants). For S-type stars, spectral types are listed as SX/Y, where X is the temperature index (not available for all sources) and Y is the abundance index. Carbon stars are listed as ``C'', as further classification was not possible. Unclassified O-rich stars are listed as ``K0-M2'' and completely unclassified stars are listed as ``U''. Stars with emission lines are denoted with ``e''. Previous identifications are listed if available from the Simbad astronomical database. Other notes: Variable stars are identified based on the presence of strong water vapour bands. Oxygen rich sources with high C/O ratios (C/O $\sim$ 0.9-0.95) are identified based on very strong H-band CO lines and no OH lines. Carbon stars with the suspected C$_2$H$_2$ feature in their H-band spectra are also noted.}\\ 
\hline 
\tiny 
No. & Name & \multicolumn{3}{c}{Photometry} & \multicolumn{2}{c}{Exposures (s)} & Spectral & Notes \\ 
 & & \multicolumn{1}{c}{r'} & \multicolumn{1}{c}{i'} & \multicolumn{1}{c}{H$\alpha$} & zJ & HK & type  \\ 
\hline 
\endfirsthead 
\caption[]{\emph{continued}}\\ 
\hline 
No. & Name & \multicolumn{3}{c}{Photometry} & \multicolumn{2}{c}{Exposures (s)} & Spectral & Notes \\ 
 & & \multicolumn{1}{c}{r'} & \multicolumn{1}{c}{i'} & \multicolumn{1}{c}{H$\alpha$} & zJ & HK & type  \\ 
\hline 
\endhead 
\hline 
\multicolumn{9}{r}{\emph{continued on next page}}\\ 
\endfoot 
\hline 
\endlastfoot 
ERSO1  & IPHAS J184857.78-021536.6 & 21.721 & 16.039 & 20.805 & 4.0 & 5.0 & M6 &  \\ 
ERSO2  & IPHAS J183432.01-011828.1 & 21.972 & 16.311 & 20.723 & 4.0 & 10.0 & M7.5 &  \\ 
ERSO3  & IPHAS J184859.24-011234.1 & 21.838 & 15.817 & 20.698 & 3.0 & 8.0 & M6 & C/O~$\sim$~0.9-0.95 \\ 
ERSO4  & IPHAS J190745.00+032227.4 & 19.366 & 15.096 & 16.747 & 10.0 & 30.0 & M4 &  \\ 
ERSO5  & IPHAS J190843.09+032752.3 & 20.200 & 15.959 & 17.802 & 20.0 & 20.0 & M3 &  \\ 
ERSO6  & IPHAS J190906.83+033654.5 & 20.503 & 15.870 & 17.981 & 15.0 & 20.0 & M4 &  \\ 
ERSO7  & IPHAS J191402.54+024348.1 & 18.055 & 14.107 & 17.529 & 3.0 & 5.0 & M9.5 & Variable \\ 
ERSO8  & IPHAS J191014.09+112409.1 & 16.667 & 12.671 & 16.043 & 3.0 & 8.0 & M6 &  \\ 
ERSO9  & IPHAS J191033.49+113644.6 & 17.325 & 13.295 & 16.476 & 8.0 & 14.0 & M6 &  \\ 
ERSO10  & IPHAS J190836.69+113729.3 & 17.370 & 13.684 & 16.912 & 4.0 & 10.0 & M6 &  \\ 
ERSO11  & IPHAS J191016.76+114134.4 & 17.202 & 13.304 & 16.737 & 8.0 & 10.0 & SX/6 &  \\ 
ERSO12  & IPHAS J190926.40+114140.0 & 18.015 & 14.107 & 17.293 & 6.0 & 20.0 & M7 &  \\ 
ERSO13  & IPHAS J194626.37+270936.0 & 17.081 & 13.541 & 16.121 & 7.0 & 20.0 & M5 &  \\ 
ERSO14  & IPHAS J185412.60-040704.6 & 17.419 & 13.724 & 16.314 & 10.0 & 15.0 & M6 &  \\ 
ERSO15  & IPHAS J193123.25+184244.7 & 15.648 & 11.972 & 14.585 & 4.0 & 8.0 & M6 &  \\ 
ERSO16  & IPHAS J000701.80+654917.9 & 16.953 & 12.850 & 15.937 & 4.0 & 10.0 & SX/5 &  \\ 
ERSO17  & IPHAS J002455.87+654955.7 & 16.815 & 13.000 & 15.836 & 3.0 & 8.0 & M2 &  \\ 
ERSO18  & IPHAS J004918.50+652822.9 & 17.141 & 13.422 & 16.115 & 10.0 & 20.0 & M3 &  \\ 
ERSO19  & IPHAS J001841.82+660645.4 & 17.366 & 13.101 & 16.192 & 5.0 & 15.0 & M8.5 & C/O~$\sim$~0.9-0.95 \\ 
ERSO20  & IPHAS J002531.86+621912.6 & 17.289 & 13.646 & 16.187 & 8.0 & 12.0 & M6.5 &  \\ 
ERSO21  & IPHAS J005934.24+651815.1 & 17.977 & 14.027 & 16.961 & 10.0 & 50.0 & M8.5 & Variable \\ 
ERSO22  & IPHAS J010743.47+630523.0 & 17.071 & 13.471 & 16.731 & 3.0 & 6.0 & SX/6 &  \\ 
ERSO23  & IPHAS J021849.42+622138.8 & 17.996 & 14.369 & 17.612 & 20.0 & 15.0 & SX/6 &  \\ 
ERSO24  & IPHAS J023951.19+555352.3 & 17.470 & 13.210 & 16.897 & 3.0 & 3.0 & SX/6 &  \\ 
ERSO25  & IPHAS J025402.88+575126.7 & 18.920 & 13.740 & 17.758 & 2.9 & 3.0 & U & Noisy~spectrum \\ 
ERSO26  & IPHAS J030552.92+542054.2 & 20.045 & 14.924 & 18.719 & 5.0 & 20.0 & M10.5 & Variable \\ 
ERSO27  & IPHAS J010744.59+590302.0 & 20.140 & 14.681 & 19.125 & 3.0 & 8.0 & SX/6e &  \\ 
ERSO28  & IPHAS J033511.00+505830.2 & 14.799 & 11.034 & 14.076 & 3.0 & 4.0 & SX/6 &  \\ 
ERSO29  & IPHAS J033938.45+521452.8 & 15.477 & 11.906 & 14.351 & 3.0 & 6.0 & M6.5 &  \\ 
ERSO30  & IPHAS J034517.16+561951.5 & 15.874 & 11.867 & 14.760 & 3.0 & 3.0 & S2/4 &  \\ 
ERSO31  & IPHAS J025540.05+602012.0 & 15.907 & 11.861 & 14.483 & 3.0 & 6.0 & S3/4 &  \\ 
ERSO32  & IPHAS J230150.16+613946.2 & 20.214 & 14.848 & 19.392 & 3.0 & 3.0 & SX/6 &  \\ 
ERSO33  & IPHAS J184927.44+034408.8 & 20.254 & 14.983 & 19.070 & 5.0 & 10.0 & SX/4 &  \\ 
ERSO34  & IPHAS J190141.34+063409.8 & 21.428 & 16.609 & 20.283 & 20.0 & 30.0 & M5 &  \\ 
ERSO35  & IPHAS J184747.70+005111.2 & 21.943 & 17.065 & 21.051 & 6.0 & 20.0 & M7.5 &  \\ 
ERSO36  & IPHAS J190032.96+030112.7 & 21.899 & 16.334 & 20.489 & 15.0 & 10.0 & M5.5 &  \\ 
ERSO37  & IPHAS J192009.24+102007.1 & 19.347 & 15.013 & 18.205 & 6.0 & 15.0 & M6 &  \\ 
ERSO38  & IPHAS J185316.59-023712.6 & 21.124 & 16.300 & 19.986 & 10.0 & 20.0 & M7 &  \\ 
ERSO39  & IPHAS J190245.00+075338.5 & 20.076 & 15.059 & 18.935 & 3.0 & 15.0 & M7.5 &  \\ 
ERSO40  & IPHAS J203602.11+380401.6 & 19.806 & 15.093 & 18.759 & 3.0 & 10.0 & SX/5 &  \\ 
ERSO41  & IPHAS J183704.11-010704.2 & 20.947 & 16.227 & 19.962 & 15.0 & 30.0 & M5 &  \\ 
ERSO42  & IPHAS J190823.19+054151.7 & 20.371 & 16.055 & 19.484 & 10.0 & 30.0 & M5 &  \\ 
ERSO43  & IPHAS J184925.41+042234.7 & 17.080 & 12.439 & 15.859 & 3.0 & 3.0 & M9.5 & Variable \\ 
ERSO44  & IPHAS J192426.56+235545.8 & 19.446 & 15.744 & 17.608 & 6.0 & 12.0 & M9 & Variable \\ 
ERSO45  & IPHAS J190943.96+134240.5 & 18.838 & 14.812 & 17.658 & 12.0 & 30.0 & M6.5 &  \\ 
ERSO46  & IPHAS J215203.82+574915.5 & 18.341 & 14.268 & 17.145 & 3.0 & 6.0 & M6.5 &  \\ 
ERSO47  & IPHAS J184336.32+010256.3 & 19.514 & 15.189 & 18.129 & 15.0 & 20.0 & K0-M2 &  \\ 
ERSO48  & IPHAS J214953.49+540713.9 & 18.512 & 14.441 & 17.256 & 8.0 & 12.0 & M5.5 &  \\ 
ERSO49  & IPHAS J202810.38+375759.8 & 18.814 & 15.170 & 17.561 & 8.0 & 15.0 & M3 &  \\ 
ERSO50  & IPHAS J190848.37+033659.1 & 19.074 & 14.946 & 17.450 & 15.0 & 30.0 & M6 &  \\ 
ERSO51  & IPHAS J214234.08+572008.9 & 16.769 & 12.543 & 15.270 & 3.0 & 3.0 & K0-M2 &  \\ 
ERSO52  & IPHAS J202503.70+371340.6 & 18.456 & 13.488 & 17.183 & 3.0 & 3.0 & S6/2 &  \\ 
ERSO53  & IPHAS J204417.37+414718.0 & 18.688 & 14.726 & 17.876 & 3.0 & 6.0 & K0-M2 &  \\ 
ERSO54  & IPHAS J200921.71+282602.7 & 19.075 & 14.625 & 18.524 & 3.0 & 10.0 & M9 & C/O~$\sim$~0.9-0.95 \\ 
ERSO55  & IPHAS J200959.88+280809.3 & 18.990 & 14.563 & 18.193 & 3.0 & 6.0 & M7 &  \\ 
ERSO56  & IPHAS J203609.30+402950.9 & 19.623 & 14.548 & 19.219 & 3.0 & 3.0 & SX/7 &  \\ 
ERSO57  & IPHAS J201448.15+361245.1 & 19.917 & 15.379 & 19.340 & 6.0 & 15.0 & K0-M2 &  \\ 
ERSO58  & IPHAS J230620.82+600432.9 & 19.028 & 13.946 & 17.877 & 3.0 & 4.0 & M8.5e &  \\ 
ERSO59  & IPHAS J211755.60+473811.1 & 19.237 & 14.023 & 17.941 & 3.0 & 8.0 & M10.5 & Variable \\ 
ERSO60  & IPHAS J195445.65+325937.9 & 20.583 & 17.011 & 19.814 & 6.0 & 40.0 & C & C$_2$H$_2$ \\ 
ERSO61  & IPHAS J231655.24+602600.6 & 17.109 & 12.382 & 15.677 & 3.0 & 6.0 & M8.5 & Variable \\ 
ERSO62  & IPHAS J005317.94+623611.5 & 20.680 & 15.204 & 19.295 & 4.0 & 15.0 & M6.5 &  \\ 
ERSO63  & IPHAS J020611.54+610528.1 & 19.627 & 14.668 & 18.603 & 4.0 & 12.0 & S4/4 &  \\ 
ERSO64  & IPHAS J230925.46+615258.6 & 21.119 & 16.220 & 19.831 & 6.0 & 15.0 & M5 &  \\ 
ERSO65  & IPHAS J011803.17+635545.3 & 19.658 & 15.178 & 18.448 & 4.0 & 12.0 & M8.5 & Variable \\ 
ERSO66  & IPHAS J011847.64+665247.8 & 18.083 & 13.528 & 16.821 & 6.0 & 15.0 & M5 &  \\ 
ERSO67  & IPHAS J204519.10+404011.4 & 17.342 & 12.815 & 15.436 & 3.0 & 8.0 & M6 &  \\ 
ERSO68  & IPHAS J202243.51+415428.2 & 21.612 & 16.500 & 19.732 & 10.0 & 20.0 & M5.5 &  \\ 
ERSO69  & IPHAS J192423.87+142621.5 & 19.897 & 15.793 & 18.540 & 30.0 & 45.0 & M5 &  \\ 
ERSO70  & IPHAS J203908.44+392129.5 & 22.680 & 17.726 & 20.723 & 30.0 & 60.0 & M4 &  \\ 
ERSO71  & IPHAS J190810.54+110315.8 & 19.830 & 15.443 & 18.388 & 30.0 & 45.0 & M4 &  \\ 
ERSO72  & IPHAS J191007.21+112222.0 & 18.918 & 14.123 & 17.375 & 5.0 & 40.0 & M7 &  \\ 
ERSO73  & IPHAS J202922.55+400537.2 & 19.790 & 16.185 & 19.876 & 12.0 & 25.0 & M2 &  \\ 
ERSO74  & IPHAS J210511.40+440531.2 & 17.208 & 13.529 & 16.693 & 25.0 & 45.0 & K0-M2 &  \\ 
ERSO75  & IPHAS J193344.25+194748.3 & 20.348 & 16.551 & 20.006 & 36.0 & 45.0 & K0-M2 &  \\ 
ERSO76  & IPHAS J211036.95+495249.2 & 18.912 & 14.887 & 18.835 & 8.0 & 40.0 & SX/7e &  \\ 
ERSO77  & IPHAS J200510.62+344753.9 & 20.691 & 16.255 & 20.125 & 20.0 & 45.0 & K0-M2 &  \\ 
ERSO78  & IPHAS J192619.66+170909.7 & 21.643 & 18.031 & 20.979 & 60.0 & 60.0 & K0-M2 &  \\ 
ERSO79  & IPHAS J192601.33+140638.6 & 21.530 & 16.683 & 20.909 & 8.0 & 25.0 & K0-M2 &  \\ 
ERSO80  & IPHAS J190708.46+044931.6 & 18.617 & 13.831 & 18.323 & 40.0 & 60.0 & SC9/8e &  \\ 
ERSO81  & IPHAS J192706.10+181527.7 & 21.212 & 17.517 & 20.917 & 40.0 & 60.0 & K0-M2 &  \\ 
ERSO82  & IPHAS J192611.47+140919.4 & 21.167 & 17.111 & 20.651 & 25.0 & 60.0 & K0-M2 &  \\ 
ERSO83  & IPHAS J190752.06+075040.4 & 20.447 & 16.704 & 19.761 & 30.0 & 60.0 & M4.5 &  \\ 
ERSO84  & IPHAS J035507.56+493357.6 & 19.512 & 14.760 & 18.285 & 3.0 & 6.0 & S5/3 &  \\ 
ERSO85  & IPHAS J041023.81+510725.2 & 17.480 & 13.694 & 16.592 & 5.0 & 25.0 & M3 &  \\ 
ERSO86  & IPHAS J041503.75+501122.9 & 15.638 & 11.831 & 14.676 & 3.0 & 8.0 & SX/5 &  \\ 
ERSO87  & IPHAS J042606.70+482016.7 & 15.800 & 12.274 & 14.777 & 3.0 & 10.0 & K0-M2 &  \\ 
ERSO88  & IPHAS J043215.53+423614.6 & 16.804 & 12.575 & 15.495 & 3.0 & 6.0 & M10e & Variable \\ 
ERSO89  & IPHAS J053653.80+311306.0 & 17.263 & 13.435 & 16.234 & 7.0 & 30.0 & K0-M2 &  \\ 
ERSO90  & IPHAS J054141.41+295318.2 & 15.779 & 11.697 & 14.697 & 3.0 & 6.0 & SX/5 &  \\ 
ERSO91  & IPHAS J054434.74+281759.5 & 15.796 & 11.795 & 14.604 & 3.0 & 3.0 & M9.5 & Variable \\ 
ERSO92  & IPHAS J054529.94+290705.2 & 16.869 & 12.034 & 15.689 & 3.0 & 3.0 & M10.5 & Variable \\ 
ERSO93  & IPHAS J054837.39+243947.0 & 16.209 & 12.659 & 15.180 & 6.0 & 15.0 & K0-M2 &  \\ 
ERSO94  & IPHAS J054921.16+264624.3 & 19.270 & 15.736 & 18.565 & 15.0 & 30.0 & K0-M2 &  \\ 
ERSO95  & IPHAS J063206.40+041718.2 & 20.305 & 16.113 & 19.430 & 30.0 & 60.0 & K0-M2 &  \\ 
ERSO96  & IPHAS J063455.83+043847.1 & 16.147 & 12.003 & 14.906 & 6.0 & 15.0 & M8.5 & Variable \\ 
ERSO97  & IPHAS J063552.51-030815.8 & 17.287 & 13.307 & 16.048 & 5.0 & 15.0 & M8 & Variable \\ 
ERSO98  & IPHAS J184029.02+035812.6 & 17.247 & 13.602 & 16.129 & 15.0 & 15.0 & M6.5 & C/O~$\sim$~0.9-0.95 \\ 
ERSO99  & IPHAS J185904.00+081851.1 & 19.728 & 15.178 & 18.575 & 20.0 & 30.0 & M6 &  \\ 
ERSO100  & IPHAS J190010.63+073538.7 & 19.472 & 15.378 & 18.669 & 15.0 & 20.0 & K0-M2e &  \\ 
ERSO101  & IPHAS J185136.91+020514.0 & 19.206 & 14.751 & 18.231 & 20.0 & 30.0 & M5 &  \\ 
ERSO102  & IPHAS J185131.52+020517.9 & 19.600 & 16.018 & 18.769 & 30.0 & 30.0 & K0-M2 &  \\ 
ERSO103  & IPHAS J192718.83+202656.9 & 18.572 & 14.828 & 17.547 & 40.0 & 60.0 & K0-M2 &  \\ 
ERSO104  & IPHAS J200348.11+290553.2 & 16.341 & 12.474 & 15.688 & 10.0 & 20.0 & SX/5 &  \\ 
ERSO105  & IPHAS J203806.06+405336.6 & 19.689 & 15.306 & 18.887 & 5.0 & 20.0 & U & C/O~$\sim$~0.9-0.95 \\ 
ERSO106  & IPHAS J203914.32+413046.4 & 20.184 & 16.329 & 19.417 & 30.0 & 60.0 & M5.5 &  \\ 
ERSO107  & IPHAS J204834.99+442726.8 & 19.590 & 15.939 & 18.883 & 25.0 & 60.0 & K0-M2 &  \\ 
ERSO108  & IPHAS J203421.64+412227.8 & 18.833 & 15.307 & 17.934 & 15.0 & 30.0 & K0-M2 &  \\ 
ERSO109  & IPHAS J192703.20+203707.8 & 17.086 & 13.250 & 15.988 & 20.0 & 20.0 & M2 &  \\ 
ERSO110  & IPHAS J185800.71+072719.4 & 18.867 & 15.135 & 17.969 & 60.0 & 60.0 & M5 &  \\ 
ERSO111  & IPHAS J203440.02+415433.0 & 18.452 & 14.670 & 17.600 & 20.0 & 30.0 & K0-M2 &  \\ 
ERSO112  & IPHAS J203511.15+403556.9 & 20.085 & 16.013 & 19.175 & 30.0 & 60.0 & K0-M2 &  \\ 
ERSO113  & IPHAS J211510.01+520218.0 & 20.735 & 16.946 & 20.096 & 20.0 & 100.0 & C &  \\ 
ERSO114  & IPHAS J184315.46+035308.8 & 18.545 & 13.805 & 17.217 & 5.0 & 30.0 & M7.5e &  \\ 
ERSO115  & IPHAS J185923.26+081024.7 & 18.414 & 13.854 & 17.286 & 4.0 & 20.0 & M6.5 &  \\ 
ERSO116  & IPHAS J185849.22+074453.2 & 18.257 & 14.481 & 17.266 & 30.0 & 30.0 & U & Noisy~spectrum \\ 
ERSO117  & IPHAS J185115.41+013909.8 & 21.098 & 16.903 & 20.396 & 12.0 & 60.0 & C & C$_2$H$_2$ \\ 
ERSO118  & IPHAS J184041.73-012350.8 & 20.621 & 16.778 & 19.608 & 60.0 & 100.0 & K0-M2 &  \\ 
ERSO119  & IPHAS J204058.42+403347.7 & 21.204 & 16.968 & 20.654 & 3.0 & 30.0 & C & C$_2$H$_2$ \\ 
ERSO120  & IPHAS J185216.79-032928.9 & 19.599 & 15.988 & 18.671 & 60.0 & 100.0 & K0-M2 &  \\ 
ERSO121  & IPHAS J212057.68+470041.4 & 20.293 & 15.932 & 20.071 & 5.0 & 20.0 & SX/6 &  \\ 
ERSO122  & IPHAS J184134.15-022446.0 & 21.844 & 16.557 & 20.254 & 60.0 & 100.0 & M6.5 &  \\ 
ERSO123  & IPHAS J194854.17+295316.1 & 17.168 & 13.561 & 16.031 & 15.0 & 30.0 & M6 &  \\ 
ERSO124  & IPHAS J202632.12+410018.8 & 18.837 & 15.148 & 18.055 & 6.0 & 60.0 & K0-M2 &  \\ 
ERSO125  & IPHAS J203219.30+404603.6 & 20.159 & 16.368 & 19.489 & 15.0 & 100.0 & K0-M2 &  \\ 
ERSO126  & IPHAS J203423.56+402354.8 & 19.999 & 16.411 & 19.267 & 15.0 & 60.0 & K0-M2 &  \\ 
ERSO127  & IPHAS J203538.75+413102.0 & 19.728 & 16.154 & 19.070 & 20.0 & 60.0 & K0-M2 &  \\ 
ERSO128  & IPHAS J203653.35+411501.2 & 18.990 & 15.425 & 18.151 & 20.0 & 60.0 & K0-M2 &  \\ 
ERSO129  & IPHAS J203232.98+404701.8 & 20.066 & 15.502 & 19.142 & 5.0 & 30.0 & M5.5 &  \\ 
ERSO130  & IPHAS J204207.60+413051.3 & 20.203 & 16.057 & 19.091 & 40.0 & 40.0 & M4 &  \\ 
ERSO131  & IPHAS J202905.52+394245.8 & 20.496 & 14.624 & 19.151 & 3.0 & 5.0 & M6.5 &  \\ 
ERSO132  & IPHAS J203250.04+414720.8 & 21.806 & 16.431 & 21.466 & 10.0 & 30.0 & SX/7 &  \\ 
ERSO133  & IPHAS J204114.29+405617.0 & 20.357 & 16.597 & 19.581 & 20.0 & 60.0 & U &  \\ 
ERSO134  & IPHAS J203053.66+403232.6 & 20.483 & 16.576 & 19.690 & 12.0 & 30.0 & M5 &  \\ 
ERSO135  & IPHAS J184226.36-021737.5 & 20.343 & 15.807 & 19.244 & 40.0 & 100.0 & M6 &  \\ 
ERSO136  & IPHAS J192459.62+170653.4 & 20.337 & 15.146 & 19.619 & 3.0 & 10.0 & C &  \\ 
ERSO137  & IPHAS J202012.43+384657.2 & 19.816 & 15.426 & 19.605 & 5.0 & 20.0 & SX/6 &  \\ 
ERSO138  & IPHAS J203741.24+412137.0 & 20.228 & 15.905 & 19.288 & 20.0 & 30.0 & M4 &  \\ 
ERSO139  & IPHAS J205032.35+413459.9 & 18.175 & 14.647 & 17.200 & 20.0 & 40.0 & K0-M2 &  \\ 
\label{observation_list} 
\end{longtable} 

\normalsize

\begin{table}
\begin{center}
\begin{tabular}{@{}lll}
\hline
No. & Name & Previous identifications \\ 
\hline
ERSO23 & IPHAS J021849.42+622138.8 & Variable~star~\citep{usat08} \\ 
ERSO24 & IPHAS J023951.19+555352.3 & EI~Per~(M8,~Mira)~\citep{bide87} \\ 
ERSO26 & IPHAS J030552.92+542054.2 & V673~Per~(Mira)~\citep{kaza03b} \\ 
ERSO27 & IPHAS J010744.59+590302.0 & V890~Cas~(Mira)~\citep{kaza03b} \\ 
ERSO43 & IPHAS J184925.41+042234.7 & Haro~Chavira~35~(M7)~\citep{skif99} \\ 
ERSO51 & IPHAS J214234.08+572008.9 & Variable~star~\citep{kun87} \\ 
ERSO58 & IPHAS J230620.82+600432.9 & QU~Cep~(M6,~Mira)~\citep{rosi76} \\ 
ERSO60 & IPHAS J195445.65+325937.9 & WC~Wolf-Rayet?~\citep{cohe95} \\ 
ERSO61 & IPHAS J231655.24+602600.6 & V563~Cas~(M6e)~\citep{rosi76} \\ 
ERSO62 & IPHAS J005317.94+623611.5 & M4~\citep{ichi81} \\ 
ERSO80 & IPHAS J190708.46+044931.6 & CSS2~30~(S-type)~\citep{step90} \\ 
ERSO91 & IPHAS J054434.74+281759.5 & IZ~Tau~(M9,~Mira)~\citep{step92} \\ 
ERSO92 & IPHAS J054529.94+290705.2 & V530~Aur~(M8,~OH/IR)~\citep{iyen98} \\ 
ERSO117 & IPHAS J185115.41+013909.8 & Carbon~star~\citep{kwok97} \\ 
ERSO119 & IPHAS J204058.42+403347.7 & Carbon~star~\citep{kwok97} \\ 
ERSO136 & IPHAS J192459.62+170653.4 & OH/IR~\citep{lesq92},~Carbon~star~\citep{kwok97} \\ 
\hline
\end{tabular}
\caption{Observed sources with previous identifications in the literature. See Section \ref{s-individual} for notes on individual sources of interest.} 
\label{appendix}
\end{center}
\end{table}

\twocolumn

The data were reduced following standard procedures for near-IR spectroscopy, using {\sc IRAF}\footnote{The Image Reduction and Analysis Facility, distributed by the National Optical Astronomy Observatory.} and the LIRIS Data Reduction\footnote{LIRISDR, developed by Jose Acosta Pulido, Instituto de Astrof'sica de Canarias.} dedicated software. Flat fields were taken at the beginning of each night with each grism and filter setup and applied to all observations as appropriate. Bad pixels were also removed at this stage of data analysis using a bad pixel mask produced from a series of short and long exposures taken at the beginning of each night.

Consecutive pairs of AB two-dimensional spectra were subtracted to remove the sky background and then co-added. Spectra were then extracted from the resulting frames and wavelength-calibrated before co-adding all the frames to provide the final spectrum. The wavelength calibration was provided by observations of argon and xenon lamps. Observations of near-IR standard stars (see Table~\ref{standards}) were made throughout each night. The choice of a slit much narrower than the seeing does not allow us to perform flux calibration using our standard star observations, and due to the intrinsic variability of the majority of these sources, would be of limited use.

\begin{table*}
\begin{center}
\footnotesize
\begin{tabular}{@{}lccccccccl}
\hline
Name & Spectral type & RA & Dec & \multicolumn{5}{c}{Magnitudes} & Nights \\
 & & (J2000) & (J2000) & B & V & J & H & K$_s$ & observed \\
\hline
HIP71819 & G2V & 14 41 28.77 & +13 36 05.3 & 8.99 & 8.40 & 7.178 & 6.933 & 6.873 & 1, 2, 3, 4, 5 \\
HIP92865 & O8V & 18 55 23.13 & +09 20 48.1 & 9.10 & 8.64 & 7.452 & 7.377 & 7.328 & 1, 4, 5 \\
HIP96037 & G2V & 19 31 37.81 & +17 46 58.6 & 8.90 & 8.26 & 6.954 & 6.588 & 6.542 & 2, 3 \\
\hline
\end{tabular}
\caption{Infrared standard stars observed each night with LIRIS on the WHT and used for data reduction. Photometric magnitudes were obtained from the Simbad astronomical database.}
\label{standards}
\end{center}
\end{table*}

Near-IR spectroscopic observations are affected by {\it telluric} absorption due to scattering and absorption by atmospheric molecules, particularly H$_2$O and CO$_2$. Telluric correction was performed by dividing science spectra by the ratio of a G2V standard star (see Table~\ref{standards}) and a rebinned solar model spectrum from \citet{kuru79}, scaled to the airmass of the observation. For those telluric features that do not only vary with airmass (such as H$_2$O) we varied the strength of this correction until an appropriate continuum could be restored. We found this method was particularly effective in correcting telluric absorption in regions with small absorption features where the underlying spectrum contained valuable information. Finally, each spectrum was divided by an estimated continuum to normalise them.

\subsection{The final data}

The resulting data set consists of 139 spectra with a resolution of 20~\AA\ in the $J$ band and and 33~\AA\ in the $H$ and $K$ bands. Using multiple observations of the O8V standard star HIP92865 (Table~\ref{standards}), we estimate a typical signal-to-noise ratio of 30. This value drops considerably in regions of strong telluric absorption. The reduced spectra are presented in a series of figures throughout this paper and will also be available in tabulated form through the VizieR service\footnote{http://vizier.u-strasbg.fr/cgi-bin/VizieR}. A number of spectra were presented by \citet{wrig08}, but we present the full sample here, continuing the numbering system used by those authors.

\section{Spectral analysis and classification}
\label{s-spectra}

In this section, we first describe the character of the LIRIS spectra we have obtained, ordering the presentation by atmospheric window ($J$ first, followed by $H$ and then $K$). When appropriate we further break down the discussion according to the three main groupings of AGB stars by photospheric pattern (O-rich stars, S-type stars and carbon stars). With this overview in place, we move on to the task of formal classification (Section~\ref{classification}).

It has been commented before \citep{wall00} that the $J$ window is often the most useful in estimating spectral type, containing features due to TiO and VO in O-rich stars, ZrO in S-type stars and C$_2$ and CN in carbon stars. The $H$ band includes two dominant sequences due to the OH and CO molecules, the former only being found in O-rich stars, as well as a strong C$_2$ feature indicative of carbon stars. The wings of the $H$ window are also particularly influenced by water vapour bands typical of the outer atmospheres of highly variable O-rich stars. The $K$ window is less distinctive, being dominated by the CO first overtone bands present in the spectra of all cool evolved stars.

Classification of the LIRIS spectra is based on a range of spectral libraries \citep[e.g.][]{joyc98, klei86, lanc00,wall00} and on discussions of the variation of near-IR molecular signatures with effective temperature and chemical abundances \citep[e.g.][]{bret90,hink89,keen80,orig93}. Where we need to extend these schemes, we will draw attention to it. Derived spectral types are listed in Table~\ref{observation_list}.

We will also draw attention to certain objects and spectral features of special interest, including those that do not fit easily within the classification scheme. We note that at the resolution of our spectra, only the strongest atomic lines are visible in all objects observed. Given that the existing classification schemes do not make use of atomic lines, we will not discuss these in detail here.

\subsection{The spectra in the $J$ window}

\subsubsection{M-type (O-rich) stars in the J-band}

\begin{figure}
\begin{center}
\includegraphics[width=370pt, angle=270]{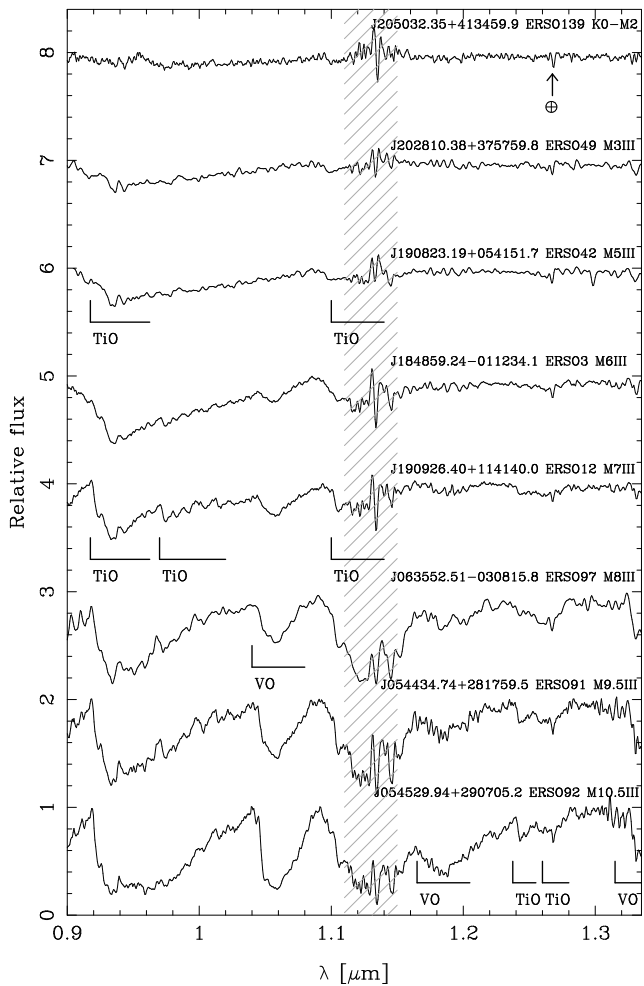}
\caption{J-band spectra of oxygen-rich evolved stars (see Table~\ref{observation_list} for details) shown in order of decreasing effective temperature. Each spectrum has been corrected for telluric absorption and has been divided by an adopted continuum. The spectra have been separated by integer values of normalised flux to make each clear and visible. The shaded area indicates a region of low atmospheric transmission where telluric correction was unable to recover a useful spectrum. Prominent molecular features have been marked, in addition to an O$_2$ telluric feature at 1.27~$\mu$m.}
\label{mtypes}
\end{center}
\end{figure}

Figure~\ref{mtypes} shows J-band spectra for a series of O-rich giant stars. The spectra contain molecular bands primarily due to TiO and, in later spectral types, VO, which identify them as O-rich stars \citep[c.f.][]{phil69,bret90}. The TiO bands appear around $\sim$3600~K ($\sim$M3) for giant stars \citep{lanc07} and get deeper as the temperature decreases. The most prominent of these is the TiO~$\epsilon$ $\Delta v = -1$ feature at 0.92~$\mu$m \citep{schi99}. This is coincident with a telluric absorption feature that we were able to successfully correct for in the majority of spectra due to its relative shallowness. For sources that are faint in this region, reconstructing the continuum via telluric correction often produced high noise levels and this feature was not always clear.

The smaller neighbouring feature at 0.97~$\mu$m is due to the TiO $\delta$ $\Delta v = -1$ transition \citep{bret90} and we find this discernible from M5-M6 onwards though it is often lost in its deeper neighbour. Other TiO lines in the J-band include the $\phi$ $\Delta v = 0$ system at 1.1~$\mu$m (just visible on the edge of a telluric H$_2$O feature) and the $\phi$ $\Delta v = -1$ system at 1.25~$\mu$m, whose two bandheads become visible from M7 onwards.

In late M-type stars the VO molecule becomes the most dominant feature, typically appearing for T$_{eff} \leq 3200$ \citep[later than M6,][]{joyc98b} and growing with decreasing temperature. The largest of these is the 0-0 head of the $A^4\Pi-X^4\Sigma^-$ $\Delta v = 0$ transition at 1.046~$\mu$m \citep{hink89}. Smaller VO features at 1.168-1.181~$\mu$m due to the $\Delta v = -1$ transition of the A-X system are also visible for very late type stars \citep{bret90}. The $\Delta v = -2$ feature is often visible at 1.325~$\mu$m, but can get lost in the deep telluric feature longwards of it \citep{hink89}.

Other features present in our spectra are a large number of atomic lines from neutral species, the strongest of which include Mn~{\sc i}~1.295~$\mu$m and Al~{\sc i}~1.313~$\mu$m. We also see the hydrogen emission lines P$\gamma$~1.093~$\mu$m and P$\beta$~1.281~$\mu$m in some of our spectra. The 3rd overtone lines of CO should lie around 1.16-1.24~$\mu$m, but \citet{hink89} searched for them in higher resolution J-band spectra and were unable to find them. We therefore rule out any of the features in this region as being due to CO and believe that the abundant neutral atomic lines predicted in this region of the spectrum by \citet{wall00} offer a more likely explanation.

\subsubsection{MS and S-type stars in the J-band}

\begin{figure}
\begin{center}
\includegraphics[width=370pt, angle=270]{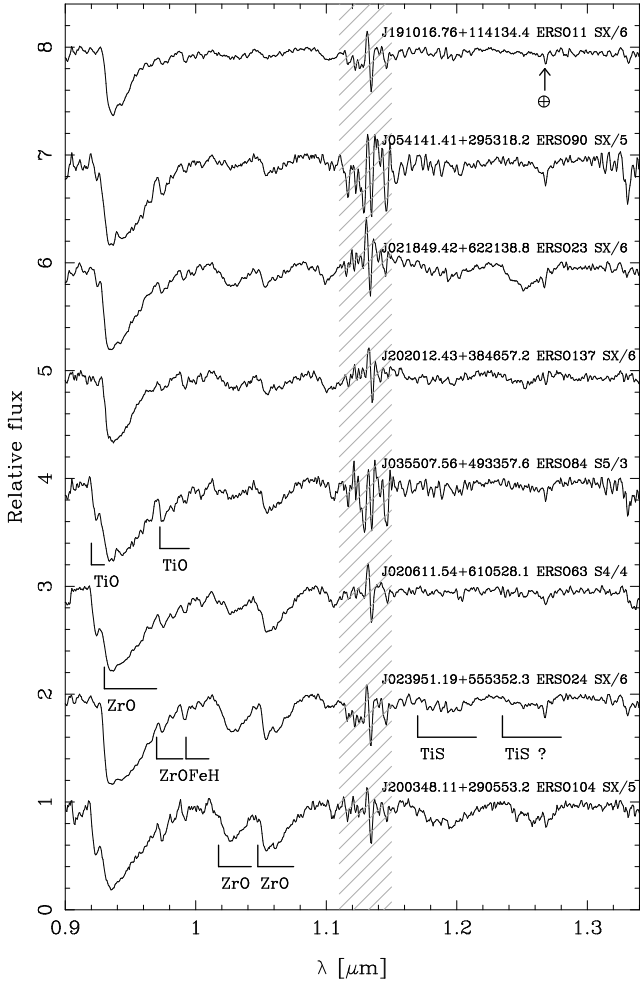}
\caption{J-band spectra of S-type evolved stars, as per Figure~\ref{mtypes} and shown in approximate order of increasing strength of the 1.0-1.1~$\mu$m ZrO features.}
\label{stypes}
\end{center}
\end{figure}

S-type giant stars are believed to be intermediate between O-rich stars (C/O $<$ 1) and C-rich stars (C/O $>$ 1) as part of an evolutionary sequence M-MS-S-SC-C (where the intermediate types show dual chemistries). This sequence is believed to be caused by an increase in the C/O ratio due to dredged-up carbon, but the abundance of Zr \citep{vant02} and variations in the photospheric temperature throughout pulsation cycles can also influence the observed spectral type \citep[e.g.][]{zijl04}. The photospheres of S-type stars exhibit near unity C/O ratios and enrichment of the s-process elements Zr, Y, La and Ce \citep{smit90}. With carbon and oxygen almost completely locked up in CO, the sulphides of these heavy elements attain an abundance equal to the oxides \citep{joyc98b}.

Figure~\ref{stypes} shows J-band spectra for a group of S-type stars which primarily feature bands due to ZrO. The strongest ZrO band visible in our spectra is at 0.93-0.96~$\mu$m \citep[the $b'^3\Pi-a^3\Delta$ system 0-0 bandhead,][]{phil79}. Its position is almost coincident with the 0.92~$\mu$m TiO feature, but the contributions from the two features may be separated and estimated. Some sources (e.g. ERSO~24) show no evidence for the TiO feature, while others (e.g. ERSO~63) show the TiO bandhead at 0.92~$\mu$m in addition to a deeper ZrO bandhead at 0.93~$\mu$m and are likely the MS-type transition objects. The form of the ZrO feature is clearly distinguishable from those due to TiO (see Figures \ref{mtypes} and \ref{stypes}).

The most unmistakable ZrO bands are the pair at 1.03 and 1.06~$\mu$m \citep[the $a^3\Delta$~0-1 and $B^1\Pi-A^1\Delta$~0-0 band heads respectively,][]{hink89}, which are easily identifiable in contrast to the VO feature found in O-rich sources. Many of these sources also show the 0.974~$\mu$m ZrO feature \citep[the head of the B-A $\Delta\nu = -1$ band,][]{davi81} and the 0.99~$\mu$m FeH band \citep[the $^4\Delta-^4\Delta$ 0-0 head,][]{lamb80}, both visible in the long-wavelength wing of the strong 0.93~$\mu$m ZrO feature. All these features have strengths dependent on the photospheric temperature, C/O ratio and heavy element abundances.

Visible in a large number of our S-type star spectra is a feature at 1.25~$\mu$m which \citet{joyc98b} observed in their spectra of S-types, but were unable to conclusively associate with a specific molecule. It coincides with the positions of the ZrS b$'$-a $\Delta\nu = 0$ sequence and the TiS A-X $\Delta\nu = 0$ sequence \citep{jons92}, both of which may contribute to the feature. 
Of the sources that display this feature we note two types. The first type shows the TiS A-X $\Delta \nu = -1$ sequence at 1.17-1.22~$\mu$m at a similar strength to the 1.25~$\mu$m feature \citep[e.g. ERSO~24, ERSO~104, or R~And,][]{jons92}, while the second type shows either a much weaker or completely absent 1.17-1.22~$\mu$m feature \citep[e.g. ERSO~23 or R~Cyg,][]{joyc98b}. 
We suggest that in the first set of cases the 1.25~$\mu$m feature originates from the TiS molecule, while in the second set of cases the feature is either purely ZrS or a combination of TiS and ZrS. This is supported by the position of the 1.25~$\mu$m feature in ERSO~23 appearing at a slightly shorter wavelength to that in ERSOs 24 or 104.

\subsubsection{Carbon stars in the J-band}

\begin{figure}
\begin{center}
\includegraphics[width=210pt, angle=270]{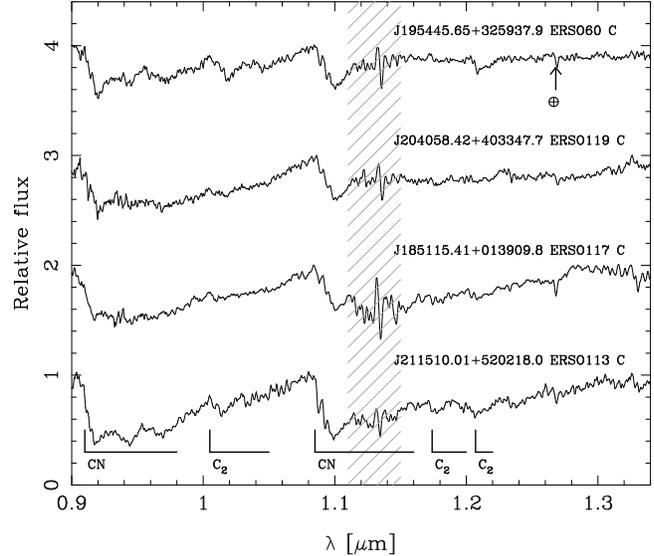}
\caption{J-band spectra of carbon stars as per Figure~\ref{mtypes} and shown in order of increasing CN feature strength.}
\label{ctypes}
\end{center}
\end{figure}

Figure~\ref{ctypes} shows J-band spectra for four carbon stars in order of increasing feature strength. The spectra of carbon stars in the J-band primarily show features due to the molecules CN and C$_2$. Clearly visible in our spectra are the CN 1-0 and 0-0 bandheads at 0.914 and 1.088~$\mu$m \citep[both part of the Red A$^2$$\Sigma$-X$^2$$\Sigma$$^+$ system,][]{joyc98b}.

Also visible in the spectra of some objects are bandheads due to the Phillips and Ballick-Ramsay systems of the C$_2$ molecule \citep{huna67,quer74}. The most prominent of these are the Phillips 1-0 transition at 1.02~$\mu$m, the Ballick Ramsay 2-0 transition at 1.174~$\mu$m and the Phillips 0-0 transition at 1.207~$\mu$m.

Carbon star spectra often show a highly "grassy" appearance which has been attributed to minor CN features and contributions from additional C-rich absorbers across these bands \citep{joyc98b}. A high fraction of the isotope $^{13}$C can also lead to many weak molecular features alongside those due to $^{12}$C.

\subsection{The spectra in the $H$ window}

Unlike the J-band, the spectral differences between M, S and C-type evolved stars in the H-band are much more subtle. The band is bordered on both sides by deep telluric H$_2$O features where transmission regularly drops to zero in our spectra. The molecular features in the H-band are principally due to the CO 2nd overtone series and the OH molecule.


\subsubsection{CO and OH bands}

\begin{figure}
\begin{center}
\includegraphics[width=330pt, angle=270]{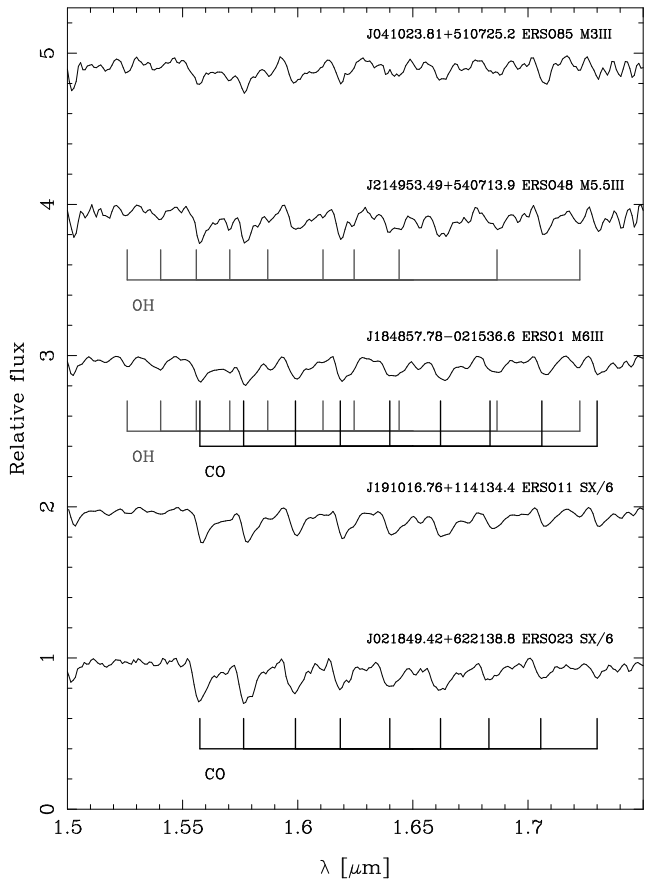}
\caption{H-band spectra of five evolved stars shown top to bottom in approximate order of C/O ratio increasing towards unity. The spectra have been divided by an adopted continuum and are separated by integer values of normalised flux. The CO and OH series are shown.}
\label{hband}
\end{center}
\end{figure}

The CO 2nd overtone ($\Delta v = -3$) series extends from 1.56~$\mu$m (3-0) to longer wavelengths with higher transition features and develops at temperatures below $\sim 5000$~K \citep{lanc07}. The strongest transition is expected to be (6-3) at 1.62~$\mu$m, with strengths decreasing before and afterwards \citep{orig93}, though at our resolution this is hard to verify. Bands from the molecular isotope $^{13}$CO are also present in a number of our H-band spectra, though the majority of their lines are confused with lines from $^{12}$CO or OH. Lines of $^{13}$CO in the H-band are very weak even for high $^{13}$C/$^{12}$C ratios because of the low optical depths of the $^{12}$CO second overtone bands (lines from the first overtone of $^{13}$CO in the K-band are easier to detect).

Some of the CO bands are blended with lines from OH, a molecule that is responsible for many features across the H-band, with the strongest at 1.537, 1.625 and 1.690~$\mu$m. As the C/O ratio increases, the strength of the OH bands gradually decreases (see sequence in Figure~\ref{hband}), while the strength of the CO bands increases. M-type stars typically show a mixture of equal strength CO and OH lines, while for S-type stars the OH lines are very weak. At C/O~$\sim$~1 the CO bands reach their peak and become well defined. For carbon stars (where C/O~$>$~1), the CO lines wane in strength since less oxygen is available (see Section~\ref{carbon-hband}). Many of the weak unidentified features in the H-band are thought to be due to CN, which produces many small features, giving the impression of a noisy looking continuum \citep{orig93}.

\subsubsection{Water vapour bands}

\begin{figure}
\begin{center}
\includegraphics[width=330pt, angle=270]{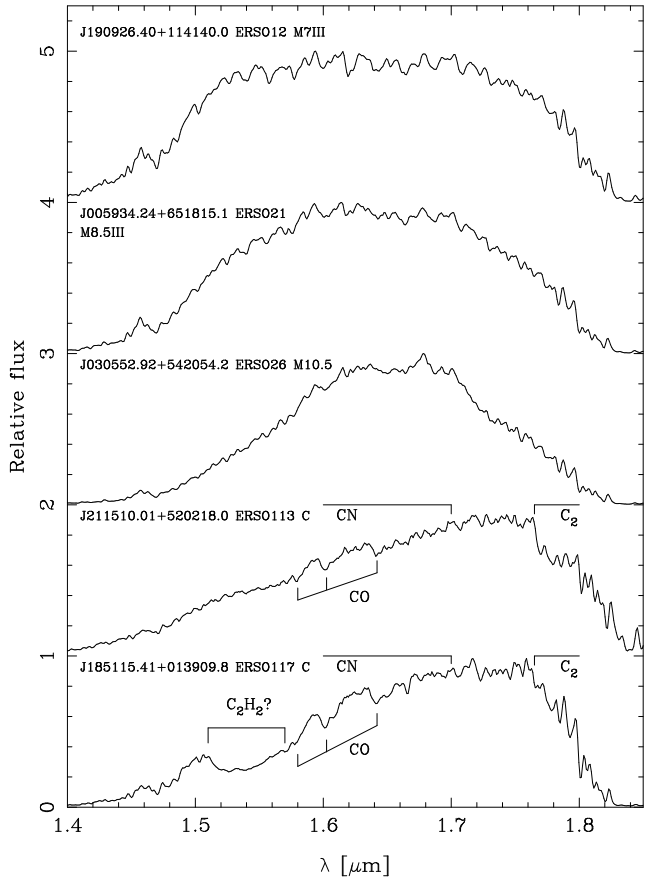}
\caption{Spectra of five stars in the H-band showing the strong effects of certain molecular features. No telluric correction has been applied to these sources and the effects of telluric absorption are visible at the short and long wavelength ends of the spectra in the ranges 1.4-1.5~$\mu$m and 1.75-1.85~$\mu$m (ERSO~12 is an example of the typical levels of telluric absorption before correction has been applied). ERSOs 12, 21, and 26 show the increasing influence of photospheric water vapour bands from negligible to very strong for very late-type O-rich sources (the absorption in the wings of ERSO~12 is believed to be due entirely to telluric water vapour and is typical of uncorrected H-band spectra). 
ERSOs 113 and 117 show the effects of C-rich molecular features on the $H$-band. CN absorption is responsible for the slope of the spectra from 1.5~$\mu$m longwards (without CN absorption the short-wavelength absorption from telluric H$_2$O would be similar to that of ERSO~12 and the full CO sequence would be visible). 
ERSO~117 also shows the unidentified feature at 1.53~$\mu$m which has been suggested to be due to C$_2$H$_2$.}
\label{hband_mult}
\end{center}
\end{figure}

The cool extended atmospheres of highly variable evolved stars enable the formation of H$_2$O, which can cause deep features coincident with the telluric absorption bands, but significantly broader \citep[e.g.][]{mats99}. Their presence implies a source with a large amplitude of variability \citep[$\delta V > 1.7$,][]{lanc00}. The exact shape of the water bands is highly dependent on the exact conditions in the outer atmosphere and varies considerably throughout the pulsation period. The observation of strong H$_2$O bands is evidence of a highly variable very late-type star, though absence of them does not disprove this. Figure~\ref{hband_mult} shows examples of the effects of water vapour bands on H-band spectra.

We find evidence for water vapour in 13 of our sources, including four with very strong absorption bands (all of which are of type M9.5 or later). The fraction of sources in our sample showing these absorption bands increases towards later spectral types, going from 57\% at M8 (4 of 7 stars) to 60\% at M9 (3 of 5 stars) and 100\% at M10 (4 stars).

\subsubsection{Carbon stars in the H-band}
\label{carbon-hband}

At C/O ratios above unity the H-band CO lines decrease in strength and the CN 0-1 bands become dominant. Because the head of this series lies in the telluric absorption feature around 1.4~$\mu$m, the band head is lost in a region of near-zero transmission. However, the series extends across most of the H-band and causes a gradient in the continuum across the band (see Figure~\ref{hband_mult}). Often the most notable feature in the near-infrared spectra of carbon stars is the C$_2$ band at 1.77~$\mu$m \citep[the Ballick-Ramsay A$'^2 \Sigma_g^-$-X$'^3 \Pi_u$~(0-0) band,][]{huna67}, which is usually the strongest of all the C$_2$ bands in the near-infrared \citep{loid01}.

Three of our five carbon stars also show an unidentified feature at 1.53~$\mu$m (see Figure~\ref{hband_mult}) that has previously been observed in cool and high Galactic latitude carbon stars \citep[e.g.][]{joyc98}.
The feature is thought to be due to the second overtone C-H stretch from molecules such as HCN or C$_2$H$_2$ and its presence is well correlated with a feature at 2.45~$\mu$m observed in laboratory spectra of C$_2$H$_2$ \citep{goeb81}. 
However \citet{joyc98} found that the 1.53~$\mu$m feature was not well correlated with a 3.1~$\mu$m feature also associated with the molecules HCN and C$_2$H$_2$. 
We calculated spectra of C$_2$H$_2$ using the HITRAN\footnote{The HIgh-resolution TRANsmission molecular absorption database.} line lists \citep{roth09} that includes lines from C$_2$H$_2$ \citep{hach02}. 
The 1.53~$\mu$m feature in our spectra matches well with the R- and P-branches of C$_2$H$_2$, confirming it as a likely carrier. 
The C$_2$H$_2$ cross section at 1.53~$\mu$m is a factor $\sim$7 larger than that of HCN, such that HCN would require a column density $\sim$7 times that of C$_2$H$_2$ to produce the same absorption. It is therefore easier to explain the observed feature with C$_2$H$_2$ unless HCN is very abundant. 
The opposite is true for the 3.1~$\mu$m feature \citep{aoki98}, which could explain the lack of correlation between the strength of this and the 1.53~$\mu$m feature observed by \citet{joyc98} if they had different dominant contributors. 

\subsection{The spectra in the $K$ window}
\label{s-kband}

\begin{figure}
\begin{center}
\includegraphics[width=330pt, angle=270]{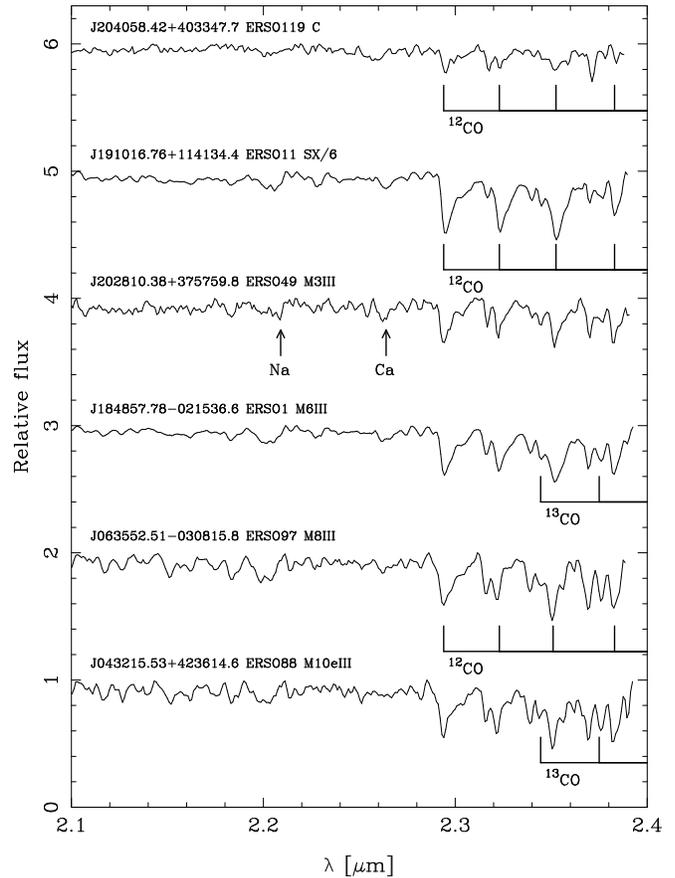}
\caption{K-band spectra of six evolved stars with different chemical types and temperature classes. Each spectrum has been divided by an adopted continuum and separated by integer values of normalised flux.}
\label{kband}
\end{center}
\end{figure}

As in the H-band the K-band is dominated by absorption from CO, though the features are stronger here. The edges of the K-band are defined by strong telluric H$_2$O features and two strong CO$_2$ features around 2~$\mu$m. The slope of the spectra in the K-band is also affected by H$_2$O absorption at the long wavelength end of the band, with many minor features contributing to the overall shape of the spectrum \citep{klei86}.

Spectra of six evolved stars in the $K$ band of different spectra types are shown in Figure~\ref{kband}. Our spectra feature four lines from the first overtone ($\Delta v = -2$) series of $^{12}$CO, starting at the 2.29~$\mu$m bandhead (2-0), with higher order terms extending to longer wavelengths. In M-type stars the strength of these lines increases with increasing luminosity and decreasing temperature and some attempts have been made to use them as an effective temperature diagnostic for giant stars \citep[e.g.][]{rami97}. However, we find only a minor correlation between spectral type and CO band strength. We observe a much stronger correlation with the C/O ratio as was also observed for the CO bands in the H-band. For M-type and carbon stars the CO bands range in strength from very weak (e.g. ERSO~119) through to moderate in strength (e.g. ERSO~1), but are always weaker than those observed for S-type stars where C/O~$\sim$~1 and the maximum amount of carbon and oxygen are available to form CO. ERSO~11 is a fine example of particularly strong CO bands.

Two lines from the first overtone series of $^{13}$CO are also visible in many of the spectra, unlike in the H-band where the second overtone series of $^{13}$CO is too weak to be visible in our spectra. The remaining features are primarily due to atomic lines from neutral species, the most prominent of which are the Ca~{\sc i} triplet at 2.265~$\mu$m and the Na~{\sc i} doublet at 2.208~$\mu$m.

\subsection{Spectral classification methods}
\label{classification}

\subsubsection{Classification of M-type stars}

O-rich sources are classified based on the strengths of the TiO and VO bands. For sources later than type M6, the VO band at 1.05~$\mu$m (e.g. Figure~\ref{stypes}) could be easily used to classify the sources. The clarity and isolation of this feature allowed clear spectral types to be assigned with an error of $\pm$1 subtype or better. Sources without VO bands, but showing the TiO bands at 0.93 or 1.10~$\mu$m were classified based on the strengths of these. Since both these features lie near to or coincident with small telluric features the error in determining spectral types for these sources is about $\pm$2 subtypes.

Sources that do not show either the TiO or VO features, yet have strong OH bands in the H-band which indicate an O-rich chemistry were harder to classify. After carefully searching for evidence of S-type features or strong CN bands \citep[indicative of M-type supergiant stars,][]{lanc00}, we were unable to assign a clear spectral type to 31 such objects. The presence of strong K-band CO lines confirms that all our sources are of late-type \citep[K0 or later,][]{lanc07}, while the lack of TiO or VO features implies they are earlier than type M2. Therefore these sources are listed as having spectral types of ``K0-M2''.

\subsubsection{Classification of S-type stars}

Sources showing ZrO features were identified as S-type stars and these spectra were then searched for the presence of O-rich or C-rich features which might identify them as MS or SC-type stars (in the 1.10~$\mu$m region small features from the molecules TiO, TiS and CN can be easily separated, allowing MS, S and SC-type stars to be identified). 
The existing classification system for S-type stars of the form SX/Y was put forward by \citet{keen80} and uses a temperature class, X (that mirrors that for O-rich M-type stars), and an abundance index, Y. The abundance index varies from 1-10 as the surface chemistry varies from O-rich (Y~$\leq 1$), to MS-type (Y~$=2-4$), S-type (Y~$= 5-6$), SC-type (Y~$=7-9$) and finally C-rich (Y~$\geq 10$) stars. 
Unfortunately the temperature class indicators used by \citet{keen80} are based on molecular bands in the optical, and the lack of any such indicators in the near-IR prevents us from assigning temperature classes to sources that do not show any O-rich features. 
We base the estimation of the abundance index on the relative strengths of the TiO, ZrO, and C$_2$ bands, as listed by \citet{keen80}. 
We also used the presence of the unidentified feature at 1.25~$\mu$m as an indicator of the abundance index since \citet{joyc98b} observed this feature only in stars with an abundance index of 5-7. 

\subsubsection{Classification of carbon stars}

Carbon stars were identified based on the presence of either the deep C$_2$ feature at 1.77~$\mu$m or deep CN bands across the J and H-bands. In the spectra of carbon stars the CN and C$_2$ features are known to increase in strength as the effective temperature decreases and the C/O ratio increases \citep{loid01}. Recent near-IR spectra of carbon stars \citep[e.g.][]{lanc00,tana07} show that these two effects are inextricably combined and that published temperature sequences or spectral class sequences do not show trends in the observed absorption features that would allow classification. Further complications in the classification of carbon stars are introduced by the different population types (R, N, or J) that are classified based on atomic and molecular features in the blue \citep[e.g.][]{keen93} and the influence of metallicity. Because of this we are unable to determine accurate spectral types for our carbon stars without fitting them to model spectra \citep[e.g.][]{tana07} that would reveal their physical characteristics. The weak CO bands observed in all our carbon stars could indicate very high C/O ratios for these sources. The trend for very late-type stars in our sample, and the potential observation of C$_2$H$_2$ in four of our five carbon stars suggests that they are likely to have particularly cool photospheres.

\section{Results of spectral classification}
\label{s-class_results}

Spectral classification was attempted for all sources observed. Of the 139 sources, 109 were found to be O-rich, 22 are of S-type and 5 are carbon stars. We were unable to classify 3 of the sources due either to unknown spectral features or noisy spectra. The assigned spectral types for these sources are listed in Table~\ref{observation_list}. Figure~\ref{colcol_liris} shows the positions of all observed sources in the IPHAS colour-colour plane, illustrating their chemical type.

\subsection{M-type stars}

\begin{figure}
\begin{center}
\includegraphics[width=188pt, angle=270]{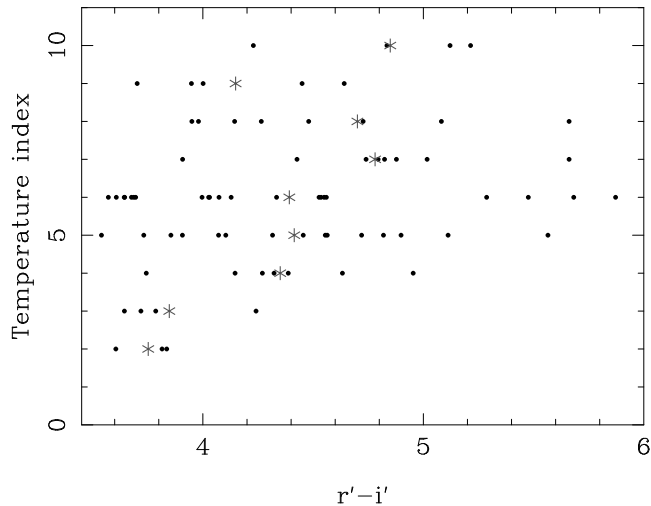}
\caption{IPHAS $(r' - i')$ colour plotted against temperature index for all observed M-type stars with spectral type of M2 or later. The stars indicate the mean colour of each spectral type. A weak trend is noted between mean colour and temperature index.}
\label{spectral_type_colour}
\end{center}
\end{figure}

There are 109 O-rich sources amongst our sample of 139 sources with LIRIS spectra. 31 of these show no evidence for TiO or VO features but do show CO bands. These have all been classified as type K0-M2. Some of these sources may be of later type but are unidentifiable because of the low transmission around 0.92~$\mu$m and the resulting inability to accurately remove the telluric absorption and identify any underlying features. \citet{wrig08} estimated that K-type giants were unlikely to significantly contribute to the ERSO region because of their bluer intrinsic colours and lower luminosities. This could suggest that these unclassified sources are of early-to-mid M-type, but that the necessary spectral features to classify them as such are unavailable.

Of the remaining 78 sources, 45 show clear VO bands indicating they are of type M6 or later. This high fraction of late-type stars supports our use of the ``extremely-red'' region of the IPHAS colour-colour diagram to select such late-type sources for our spectral library. The M-type stars show an even distribution across the colour-colour plane in Figure~\ref{colcol_liris}. Figure~\ref{spectral_type_colour} shows the $(r' - i')$ colours of all observed M-type stars with spectral types of M2 or later (earlier types are not included because of the large errors associated with determining spectral types from their relatively featureless spectra). A weak trend between $(r' - i')$ colour and spectral type does exist, though the spread is large and influenced by small number statistics. It is worth noting that the reddest source at each spectral type increases almost linearly from M2 to M6. This is most likely due to a combination of intrinsic colour and the fact that sources of later spectral type will be more luminous (therefore likely to be more distant and experience greater interstellar reddening) and may have undergone more mass loss (therefore have greater circumstellar reddening).

\subsubsection{Emission-line sources}

We searched for emission-lines in all our spectra, particularly for lines from the Paschen series found in the near-IR, e.g. P$\beta$ at 1.282~$\mu$m and P$\gamma$ at 1.094~$\mu$m. Mira variables are known to exhibit phase-dependent emission-lines in the spectra due to shocks in their atmospheres from stellar pulsations \citep[e.g.][]{hink79}. We observe emission lines in six of our spectra, three of which are oxygen-rich and three are S-type stars. Of the four stars with emission lines and for which temperature class information is available, all are of type M7 (or equivalent) or later suggesting that later type stars experience more shocks in their atmospheres. Since these shocks are thought to be due to stellar pulsations, this supports the belief that pulsations are stronger and more regular in stars of later spectral-type. The $(r' - $H$\alpha)$ colours of all these sources are not particularly large, supporting the suggestion by \citet{wrig08} that the 2~yr separation between IPHAS photometric measurements and these spectroscopic observations is longer than the temporary nature of these spectral features \citep[likely to be less than the pulsation period of the star,][]{bess96}.

\begin{figure}
\begin{center}
\includegraphics[width=188pt, angle=270]{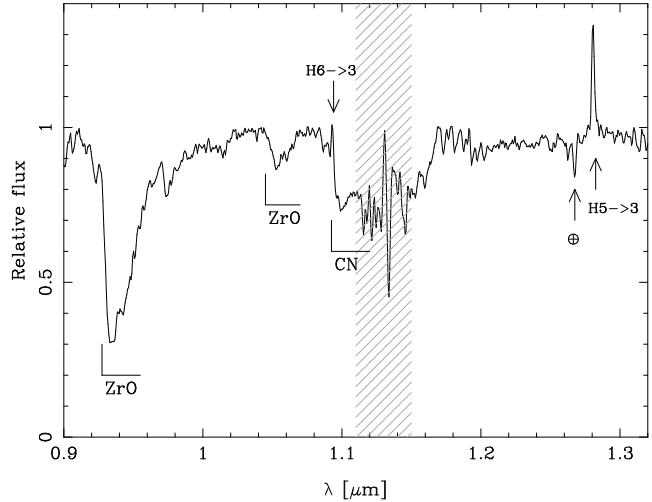}
\caption{LIRIS $J$ band spectrum of ERSO~27. The spectrum has been corrected for telluric absorption (affected region marked) and divided by an adopted continuum. Molecular features typical of S-type stars are shown, as are the Paschen $\beta$ and $\gamma$ emission lines.}
\label{erso27}
\end{center}
\end{figure}

The object with the strongest Paschen $\beta$ emission line is ERSO~27 (V890~Cas) which has $(r' - $H$\alpha) = 1.02$ and whose spectrum is shown in Figure~\ref{erso27}. While it does not have a particularly high $(r' - $H$\alpha)$ colour, it is an S-type star, which were shown by \citet{wrig08} to have lower $(r' - $H$\alpha)$ colours compared to O-rich stars, and so the effects of this and an H$\alpha$ emission line may be cancelling each other out.

\subsection{S-type and carbon stars}
\label{types-stypes}

We find 22 S-type stars and 5 carbon stars in our spectral library. More than half of the S-type stars show no evidence for O-rich features, which gives them abundance indices of 6 or more \citep{keen80}. 
We find only 1 S-type star with an abundance index $\leq 2$, which we attribute to the difficulty in identifying the weak S-type features in these sources against the stronger O-rich features. A small number of O-rich stars may therefore have been mis-classified as such. 
There is a notable concentration of S-type stars towards lower $(r' - $H$\alpha$) colours, supporting the trend originally noted by \citet{wrig08}, while the carbon stars show intermediate colours between the O-rich and S-type stars \citep[an effect commented on by][]{drew05}. The reason for these colours is that in O-rich stars TiO features cause a falsely low continuum in the $r'$ band, while the H$\alpha$ filter excludes these features resulting in a large ($r' - $H$\alpha$) colour. S-type stars however have weaker ZrO features across the $r'$ filter and a ZrO feature coincident with the H$\alpha$ filter, causing absorption resulting in a much lower ($r' - $H$\alpha$) colour \citep[see Figure~19 of][for spectral examples of this effect]{wrig08}.

\begin{figure}
\begin{center}
\includegraphics[width=188pt, angle=270]{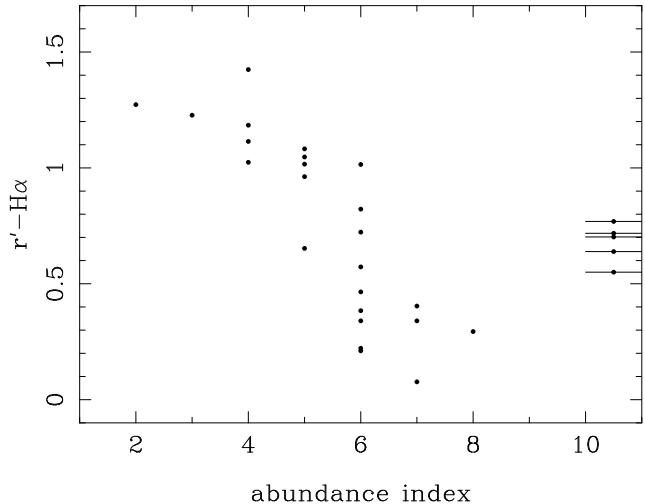}
\caption{IPHAS $(r' - $H$\alpha)$ colour plotted against the \citet{keen80} abundance index for all carbon and S-type stars. Though carbon stars are not fully included on the scale of \citet{keen80}, we have included them here with an abundance index $\geq$10 to represent the chemical evolution of an evolved star. O-rich stars have an abundance indices of $\leq 1$ and are not shown on this figure. The trend of increasing $(r' - $H$\alpha)$ colour with increasing $(r' - i')$ colour is also not considered in this figure.}
\label{abundance_index}
\end{center}
\end{figure}

To illustrate this effect we show in Figure~\ref{abundance_index} the ($r' - $H$\alpha$) colour as a function of the abundance index used to classify S-type stars. Despite the small number of sources we note a clear trend for stars with a higher abundance index to have smaller ($r' - $H$\alpha$) colours. We explain this trend as due to the waning strength of the TiO features that give rise to high ($r' - $H$\alpha$) colours and the growing strength of ZrO features that give rise to low ($r' - $H$\alpha$) colours. 
The minimum ($r' - $H$\alpha$) colour appears to occur at an abundance index of 6~--~8, the stage where ZrO-dominated spectra shift to those dominated by the sodium D lines. 
It should be noted however that this relationship is only based on sources in the ERSO region of the IPHAS colour-colour plane and does not consider less reddened sources. It also does not consider any influence on the ($r' - $H$\alpha$) colour of sources based on their ($r' - i'$) colour, for which the giant branch shows a clear gradient \citep[e.g.][]{drew05,wrig08}. O-rich sources with lower $(r' - i')$ colours will have lower ($r' - $H$\alpha)$ colours and therefore a more accurate relationship on the $(r' - $H$\alpha$) deficit must be determined as a function of the $(r' - i')$ colour.

The abundance index was considered by \citet{keen80} to be a good indicator of the C/O ratio. However, the ZrO and TiO bands that determine the abundance index are dependent on the Zr/Ti ratio and the photospheric temperature as well \citep[e.g.][]{zijl04,garc07}. It had been thought that the Zr/Ti ratio might scale with the C/O ratio \citep{scal76}, because both carbon and $s$-process elements are dredged to the surface during mixing events on the AGB. However a direct relationship has not been conclusively established \citep{vant02} and the two mixing rates may be different or the ratios start from different initial values. 
If this were the case two scenarios would arise where C/O exceeds unity either after or before the Zr/Ti ratio reaches the level where ZrO bands outweigh TiO bands. In the former situation we would expect to see evolution follow the M-S-C sequence, with the S-type phase lasting until C/O exceeds unity. In the latter case the star might evolve straight from O-rich to C-rich without an S-type phase, potentially passing through a phase where all the C and O is tied up in CO and only sulphides are visible. A source such as ERSO105 with strong CO bands and no identifiable O-rich, S-type or C-rich features might be an example of such an object.

\subsubsection{Extrinsic and intrinsic S-type stars}

A small subset of S-type stars (now dubbed {\it extrinsic} S stars) are believed to acquire their Zr not from the third dredge-up (as is typical for AGB stars) but from mass-transfer from an evolved binary companion during an earlier evolutionary phase \citep{jori88}. When the star later evolved onto the red giant branch (RGB), ZrO was able to form. Extrinsic and intrinsic (true AGB stars) S-type stars were originally separated by studying lines from the unstable element Tc that are observed in the atmospheres of AGB stars (following nucleosynthesis and the third dredge-up) but not in the less-evolved RGB stars where it has decayed \citep{brow90}.
To determine if the S-type stars identified in this work are intrinsic or extrinsic, we have utilised the near- and mid-infrared colour-colour diagrams presented by \citet{yang06}. They found that the $(K - \mathrm{[12]})$ vs $(J - \mathrm{[25]})$ colour-colour diagram could be used to separate the two types of stars, approximately separating the RGB stars with little or no circumstellar material from the more evolved AGB stars with considerable circumstellar material. We combined 2MASS \citep[Two-Micron All Sky Survey,][]{skru06} near-IR photometry with either IRAS \citep[Infrared Astronomical Satellite,][]{neug84} or MSX \citep[Midcourse Space Experiment,][]{egan96} mid-IR fluxes as per \citet{wrig08}. We find suitable associations for 18 of the 22 S-type stars and that all have colours placing them in the region of the near- and mid-IR colour-colour diagram characterised by intrinsic S-type stars. It is likely that our colour selection criteria for these spectra has preferentially selected the redder and more luminous AGB stars over red giant branch stars and has therefore selected intrinsic S-type stars over extrinsic types.

\subsection{Notes on individual sources}
\label{s-individual}

The S-type star ERSO~24 lies 8.6\arcs\ from the variable star EI~Per, which \citet{bide87} identified from infrared plates as a Mira of spectral type M8. Our spectra show very strong ZrO bands (see Figure~\ref{stypes}) and we find no evidence for O-rich features, including the H-band OH lines. IPHAS images of ERSO~24 show no other sources within 30\arcs\ with reddened colours. Despite the difference in spectral type between EI~Per and ERSO~24 we find no other candidates for EI~Per in the IPHAS images so assume that the two are the same star. The star has either undergone a recent abundance change from O-rich to S-type or the previous unverifiable spectral type identification was inaccurate.

ERSO~60 is located 6.4\arcs\ from an IRAS source that \citet{cohe95} suggested could be a Wolf-Rayet candidate based on its position in the IRAS two colour diagram. An inspection of IPHAS images reveal no other highly reddened sources in the vicinity, suggesting that the IRAS source is most likely associated with ERSO~60, the spectra of which indicate it is certainly a C-rich AGB star.

ERSO~80 was classified by \citet{step90} as an S-type star, who noted that the object had no TiO bands, weak LaO bands and was quite red. Our spectra do not show any TiO or ZrO features, but do show moderate CN bands at 1.088~$\mu$m, as well as a very clean $H$-band CO spectrum with no OH bands, both indicative of a near unity C/O ratio. There is no evidence for any bands of C$_2$, which can appear for C/O$ > 1$, so our spectral classification is limited to an abundance class of 8, which \citet{keen80} define as a star having no ZrO or C$_2$, and C/O~$\sim 1$. The deep CO bands and the presence of LaO, which is only thought to be visible for T$_{eff} < 2800$~K, indicates a relatively cool photosphere and a temperature class around $9 \pm 2$. Therefore we assign ERSO~80 a spectral classification of S9/8 on the S-type classification system, or SC9/8 since it is an SC-type star.

ERSO~136 is listed in the General catalog of galactic carbon stars \citep{alks01}, but its position in the IRAS colour-colour diagram suggests O-rich circumstellar chemistry, as noted by \citet{chen03} who suggested the star is actually O-rich. IRAS low-resolution spectra \citep{omon93} show no evidence for silicate dust features and our spectra contain weak CN lines in the J- and H-bands, which indicates a potentially C-rich object. The lack of both the 1.77~$\mu$m C$_2$ band and very strong CO bands indicates that the C/O ratio is near unity. If this source has only recently become C-rich it may explain its ambiguous mid- and far-infrared colours and spectra.

\section{Conclusions}

We have presented near-IR spectra of 139 AGB stars that have been classified by comparison with existing spectral libraries. Our spectral library covers the full range of O-rich, S-type and carbon stars, with a significant number of very late-type sources in all chemical types. This was achieved by selecting sources from the extremely red region of the IPHAS colour-colour plane, which has been shown to be dominated by late-type AGB stars. Classification of late K-type and early M-type stars was not possible due to the lack of clear classification diagnostics at our resolution for these sources.

The spectral library also includes a significant fraction of S-type stars, which have been classified by temperature index and abundance index where possible. 
We find a strong correlation between the IPHAS $(r' - $H$\alpha)$ colour and the C/O abundance index for S-type and carbon stars. Combined with photometry in the near-IR \citep[e.g.][]{cion03}, this relation could be used to separate O-rich, S-type and carbon stars based on photometry alone. Given the recent generation of optical and near-IR photometric surveys across the Galactic Plane, data are available to provide a broad identification of AGB-star chemistries across much of the Galaxy from the solar neighbourhood outward. Since the fraction of O-rich, S-type and carbon stars is a known indicator of metallicity, this could be used to trace the metallicity gradient in the outer Galactic disk. Further spectra will be necessary to refine this relationship and determine its dependence on the IPHAS $(r' - i')$ colour.


\section{Acknowledgments}

This work is based in part on observations made with the Isaac Newton Telescope and the William Herschel Telescope, operated on the island of La Palma by the Isaac Newton Group. Observations on the WHT were obtained through an International Time Programme, awarded to the IPHAS collaboration. 
It also partly makes use of data products from 2MASS, IRAS and MSX, which are jointly run by the Infrared Processing and Analysis Center and the California Institute of Technology, and funded by the National Aeronautics and Space Administration and the National Science Foundation.  This research has made use of the Simbad database, operated at CDS, Strasbourg, France, and {\sc IRAF}, operated by the Association of Universities for the Research in Astronomy, Inc., under cooperative agreement with the National Science Foundation. NJW was supported by a PPARC Studentship and a Smithsonian Astrophysical Observatory Pre-doctoral Fellowship.

\bibliographystyle{mn2e}
\bibliography{/Users/nick/Documents/Work/tex_papers/bibliography.bib}
\bsp

\label{lastpage}

\end{document}